\documentclass[a4paper,12pt]{article}

\usepackage{epsfig}
\begin{document}
\large
\vskip 1cm
\section*{The deterministic-stochastic flow model}
\vskip 1cm
{\it A.P. Buslaev, V.M. Prikhodko, 
A.G. Tatashev, M.V.Yashina
}
\vskip 0.5cm  
{
The Moscow  State   Automobile \& Road  Technical    
University,} 

{\it
\centerline{E-mail: Busl@math.madi.ru}}

\subsection*{1. The basic conceptions}
Let us consider the movement of particles (vehicles) on multilane road
fragment. Let $v$ be a velocity of regular movement determined by a
number of particles called {\it slow vehicles} that differs 
sufficiently from zero. Let $d=d(v)$ be a {\it dynamic distance}, that 
includes one lane part of a
road with the length that covers the length of vehicle and braking way (the size
of discretization step by space coordinate) [1], fig. 1.

\begin{figure}[h]
\centerline{\includegraphics[width=12cm]{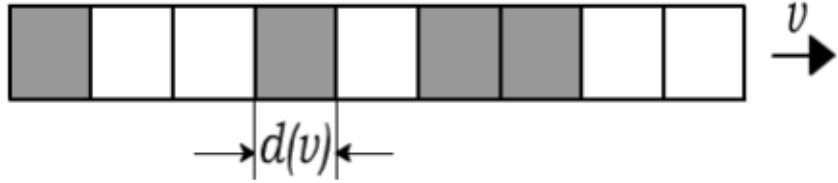}}
\caption{Discretization}
\label{}
\end{figure}

The dependence $d(v)$ can be approximated by the quadratic relation
$$d(v)=c_0+c_1 v+c_2 v^2$$
where $c_0$ is the length of vehicle base, $c_1$ is a coefficient which
is connected with regard of driver's reaction in case of
unexpected traffic condition change, $c_2v^2$ is an evaluation of braking
way. For example let us consider Tanaka model, [2]
$$d(v) = 5.7+0.14v+0.0022 v^2, \eqno(1)$$
$d(v)(m),$ $v$ (km/h). If in the equation (1)  the  velocity
is measured in m/sec, then $c_0 =5.7$ m, $c_1= 0.14\cdot 3.6= 0.504$ sec,
$c_2=0.0022\cdot (3.6)^2 =0.0285$ $\rm{sec^2/m}.$ In this case we have
$$d(v)=5.7+0.504v+0.0285v^2.\eqno(2)$$ 

The coefficient $c_2$ depends on condition of the road covering. 
So, according [3], for the wet asphalt-concrete road covering the coefficient $c_2$                     
is in two times larger than for dry asphalt-concrete road covering,
that is $d(v)=5.7+0.504v+0.057v^2,$ and for a road covered with ice
$d(v)=5.7+0.504v+0.165v^2.$ Fig. 2 shows the dependence 
$d(v)$ for different conditions of road covering. The graph 
of the function $d(v)$ for different conditions of the road 
covering is shown on fig. 2.

\begin{figure}[h]
\centerline{\includegraphics[width=12cm]{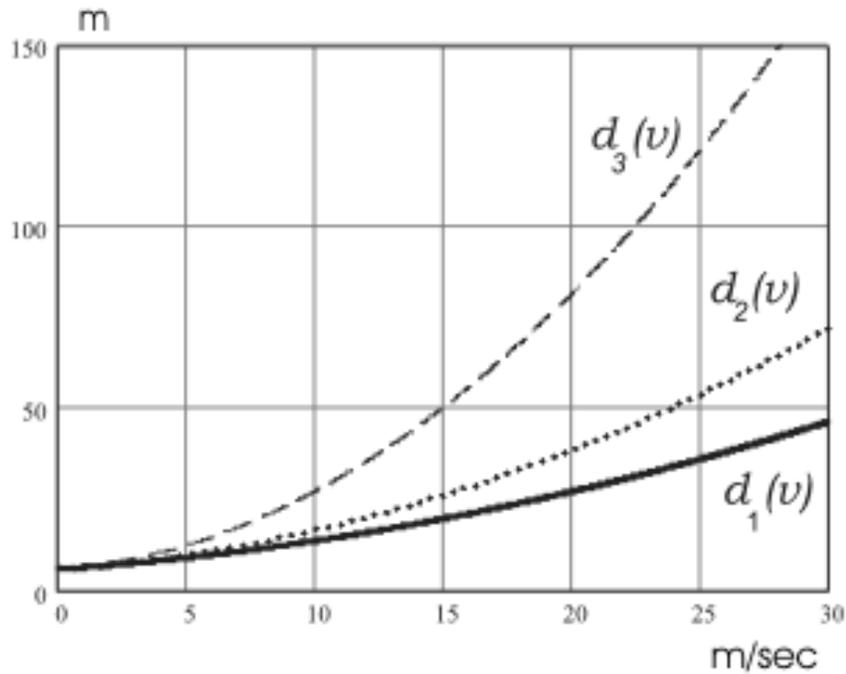}}
\caption{Dynamic distance: $d_1(v)$~--- dry asphalt; 
$d_2(v)$~--- wet asphalt; $d_3(v)$~--- road covered with ice
}
\label{}
\end{figure}

Thus the immediate location of a vehicle on the road can be 
shown by the cell field, which for simplicity is synchronized
in relation to lanes (fig. 3).

\begin{figure}[h]
\centerline{\includegraphics[width=12cm]{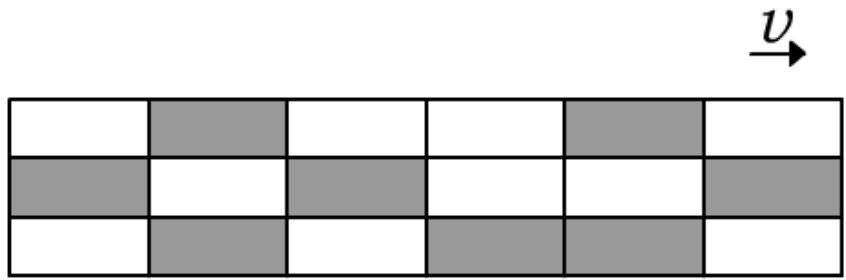}}
\caption{The three-lane cell field
}
\label{}
\end{figure}

Let $T$ be magnitude of discrete time unit, that fixes flow state. If the
"snap" does not change then the flow is supposed to be {\it steady. The ratio
of the number of occupied cells to their total number is called the
regularity, $r\in [0;1]$.} If $r=1$ then we have the steady flow, an 
{\it army column,} which presents the movement of the column with the
constant velocity $v=const.$ It is a consequence of busy cells. If $r < 1,$ 
then the individual transition of a vehicle to a front or neighbour cell 
diagonally is possible in a time unit $T$ (fig. 4).

\begin{figure}[h]
\centerline{\includegraphics[width=9cm]{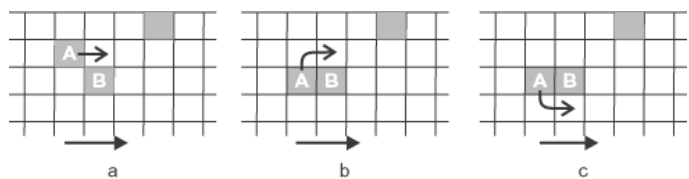}}
\caption{
Manoeuvres on the road
}
\label{}
\end{figure}

These transitions are caused by several reasons but one of the most
important is that drivers want to drive their cars with a higher velocity. Let
$p= p(t)$ be {\it a stochastic measure of individual transitions at time
unit $t$ to the cell ahead when this cell is empty.} Of course this
value also depends on other characteristics of the flow and this is the
subject of the further consideration.

Thus {\it each car will do an attempt to move forward with probability
$p(T)$ independent of the behavior of the other vehicles.} If $r \approx 0$
then there is no obstacle for such transitions as a rule, and if 
$r$ is essentially greater than 0 then the considered measure depends on $r.$
Let $p(r,t)$ be this measure. It is clear that $p(0,t)=p(T)$ and $p(r,t)$
is an non-increasing function $r\in [0;1],$  $p(1,T)=0.$ 

Let us evaluate the function $p(r,t).$ Suppose that {\it in the neighbourhood 
of the considered cell the states of the three neighbour cells that follow 
ahead are independent (fig. 5).}

\begin{figure}[h]
\centerline{\includegraphics[width=9cm]{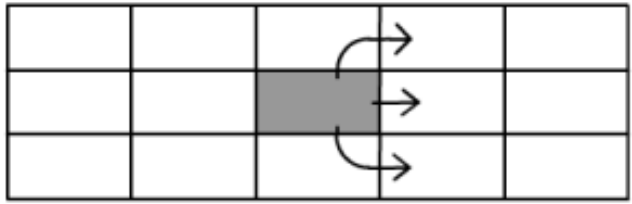}}
\caption{
Manoeuvres from an inner cell
}
\label{}
\end{figure}

Thus accurate to the coefficient $p(T)$ the probability $p_i(r;T)$ of a
transition equals
$$R_3(r)=(1-r)\cdot 1+r(2(1-r)^2-(1-r)^4)=$$
$$=(1-r)(1+2r(1-r)-r(1-r)^3)=$$
$$=(1-r)(1+2r-r^2-r+3r^2-3r^3+r^4)=$$
$$=(1-r)(1+r+r^2-3r^3+r^4)=$$
$$=1-4r^3+4r^4-r^5.$$
Similarly, for the case of inside lane on a multilane
road (or on a two-lane road), we have
$$R_2(r)=1-2r^2+r^3.$$

At last for the movement on a lane (or for {\it canalized movement}) we
obtain
$$R_1(r)=1-r.$$
Thus
$$p_1(r,T)\simeq p(T)(1-r).$$
$$p_2(r,T)\simeq p(T)(1-2r^2+r^3);$$
$$p_3(r,T)\simeq p(T)(1-4r^3+4r^4-r^5),$$
(fig. 6).

\smallskip
\begin{figure}[h]
\centerline{\includegraphics[width=9cm]{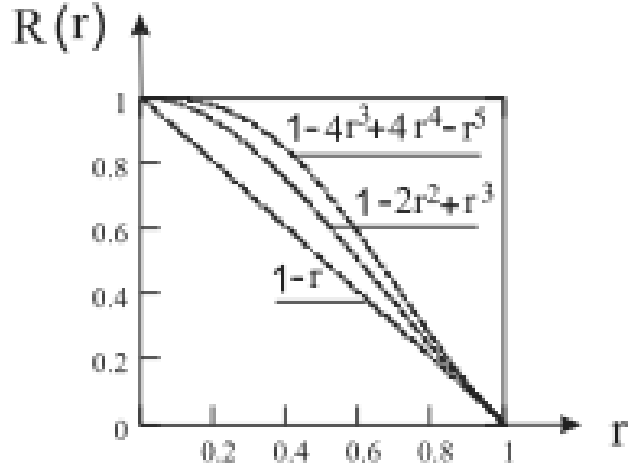}}
\caption{
Evaluation $R_i(r)$ of dependence $p_i(r;T)/p(T)$}
\label{}
\end{figure}

In common case it can be considered that $p(r;T)$ is a continuous
function on $r;$ (the addition of one busy cell cannot have any
essential influence upon the mean  velocity).

At last in common case the considered function can be also depend on $v.$
Really, every driver has his own knowledge of velocity. Therefore the additive
component to expected velocity ($v_1 > v)$ is compensated by individual
attempts. For example according to the next scheme we have
$v+d(v)p(r,T,v)/T=v_1.$

\section*{2. DST-flows (Deterministic-stochastic traffic)}

{\it The regular velocity can be also considered as the determinate component
$v_{det}$ of the flow.} Beside the collective motive of the particle
behavior in the flow each particle also has its own intensions. Hence the
flow can be presented as a composition of {\it total} (common, socialist)
and {\it private} (liberal, individual) behavior. Let us consider a flow
which consists of the particles with identical strategies of behavior
{\it (homogeneous  dst-flow},). Then the flow velocity is the sum of the
deterministic and stochastic components
$$v_{dst} =v_{det} +v_{st}\eqno(3)$$
where the stochastic transitions are independent and
distributed equally. In this relation average meaning of individual
transition is
$${\bar v}_{st} = p(r,T,v)\frac{d(v)}{T}.$$

Hence {\it the average value and the dispersion of the velocity} is
accordingly equal to
$$\bar v = v+p(r,T,v)\frac{d(v)}{T},\eqno(4)$$
$$\bar{ v^2} =p(r,T,v)(1- p(r,T,v))(\frac{d(v)}{T})^2.$$
Let $\rho$ be the density of traffic flow per lane. Then
$$\bar{\rho}=\frac{r}{d(v)}.\eqno(5)$$
Hence we have intensity per lane
$$ \bar{q} \simeq
 \rho \times \bar {v} = 
\frac{r}{d(v)}(v+ p(r,T,v)\frac{d(v)}{T})=$$
$$ = \frac{rv}{d(v)} + \frac{r p(r,T,v)}{T} .
\eqno (6)$$

The function (6) depends on three variables. As
$$p_i(r,T,v)\simeq p(0,T,v)R_i(r)$$
we have ($i$ is number of lanes)
$$\bar q_i=\frac{rv}{d(v)}+\frac{rp(0,T,v)R_i(r)}{T}.$$

Suppose $T \to 0.$ Reassume that
$$\frac{p(0,T,v)}{T}\to p(v).\eqno(7)$$
Then
$${\bar v}_{st}(i)=p(0,T,v)R_i(r)\frac{d(v)}{T}=p(v)R_i(r)d(v)$$
and the intensity is equal to
$${\bar q}_i(v,p(v),r)=\frac{rv}{d(v)}+rp(v)R_i(r).\eqno(8)$$

According to (5) we receive
$${\bar q}_i(\rho,v,p(v),\rho)=\rho v+\rho d(v)p(v)R_i(\rho d(v)).\eqno(9)$$

{\it Equation (9) generalizes the classical relation (main diagram) which is
obtained for $p(v)=0.$}

 In this case
$${\bar v}_i=v+p(v)R_i(r)d(v).$$

\section*{3. The single-lane traffic "Regularity-Velocity-Rate"}

Let us consider the case of one lane, $n=1.$
We have ${\bar v}_{st}=p(v)(1-r)d(v)$
and the equation (6) for small $T$ can be written as
$${\bar q}_1\simeq\frac{rv}{d(v)}+p(v)r(1-r)=v\rho+p(v)d(v)\rho(1-d(v)\rho),$$
where $\rho d(v)=r\le 1.$ For $\rho d(v)>1$ the dynamic distance is
not regarded and for this reason $v$ diminishes. 

As $d(v)=\frac{r}{\rho}=c_0+c_1v+c_2v^2$ we have
$${\bar q}_1\simeq{\bar q}_1(r,v,p)=\frac{rv}{c_0+c_1 v+
c_2 v^2}+pr(1-r),\eqno(10)$$
where coefficients $c_0,$ $c_1,$ $c_2$ are assigned as in 
equation (1); $p=p(v)$ 1/sec.

The function $\bar{q}_1$ is defined on the rectangular
$0<r<1,$ $0<v<v_{max}.$ Suppose $p\equiv 1.$ Let us represent the graph of the
intensity function (fig.7). 

For the same values of parameters the dependence of the individual 
velocity on $r$ and $v$ is shown on fig. 8.

\smallskip
\begin{figure}[h]
\centerline{\includegraphics[width=9cm]{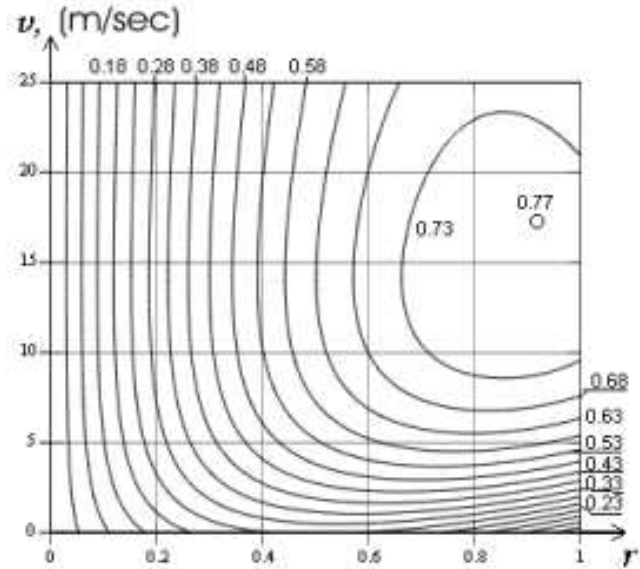}}
\caption{
Function $ {\bar q}_1(r,v,1)$ vehicle/sec}
\label{}
\end{figure}

\smallskip
\begin{figure}[h]
\centerline{\includegraphics[width=9cm]{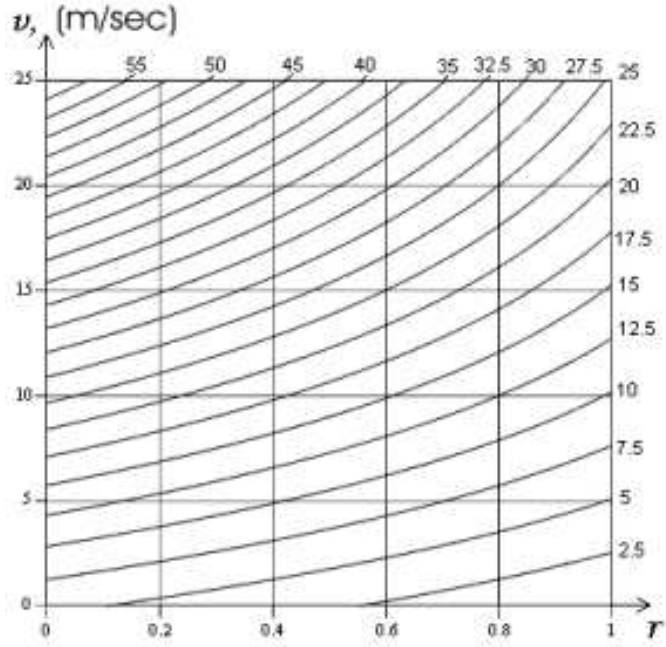}}
\caption{
Function $ {\bar v}_1(r,v,1)$ m/sec}
\label{}
\end{figure}

It can be seen from fig. 7 that the largest flow rate is reached when the
values of basic variables $r$ and $v$ are near by (0,8;15). In some
neighborhood of point of maximum the trajectories of intensity 
level lines are close of and are included to the considered set 
of values. Thus {\it the flow intensity can be still invariant 
when the parameters of the flow change.} 

Comparing the behavior of graphs of the intensity and the velocity
(fig. 7--8) we note that {\it the conflict 
of the collective and individual purposes occurs.} 
Just, if the velocity remains constant then the intensity
changes, as the level lines intersect.

In common case we have
$$\frac{\partial q_1}{\partial{r}}=\frac{v}{c_0+c_1 v+c_2 v^2}+p-2rp=0,$$
$$\frac{\partial q_1}{\partial{v}}=\frac{r}{c_0+c_1 v+c_2 v^2}-
\frac{rv(c_1+c_2v)}{(c_0+c_1v+c_2 v^2)^2}=0,$$
i.e. $c_0+c_1 v+c-2v^2=v(c_1+2c_2v)$ and $c_2v^2=c_0$. Thus
$$v^*=\sqrt{\frac{c_0}{c_2}},$$
$$r^*=\frac{1}{2}+\frac{1}{2p}\frac{v^*}{c_0+c_1v^*+c_2v^{*2}}.$$

\subsection*{4. Stability of the traffic and scattering
of the fundamental diagram}

\smallskip
\begin{figure}[h]
\centerline{\includegraphics[width=9cm]{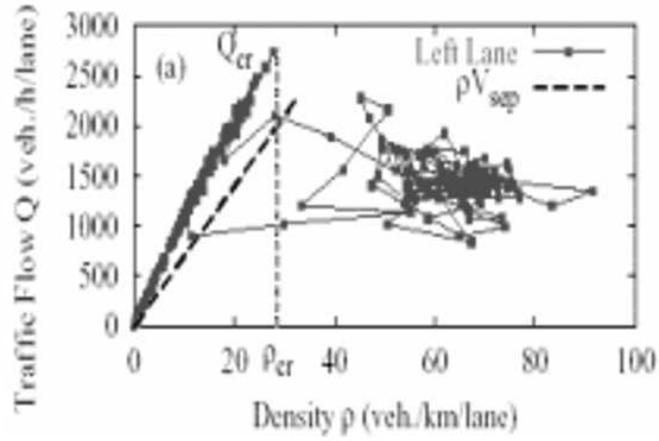}}
\caption{
Experimental data in field "density--intensity", [4]
}
\label{}
\end{figure}

\newpage
\begin{center}
\epsfig{figure=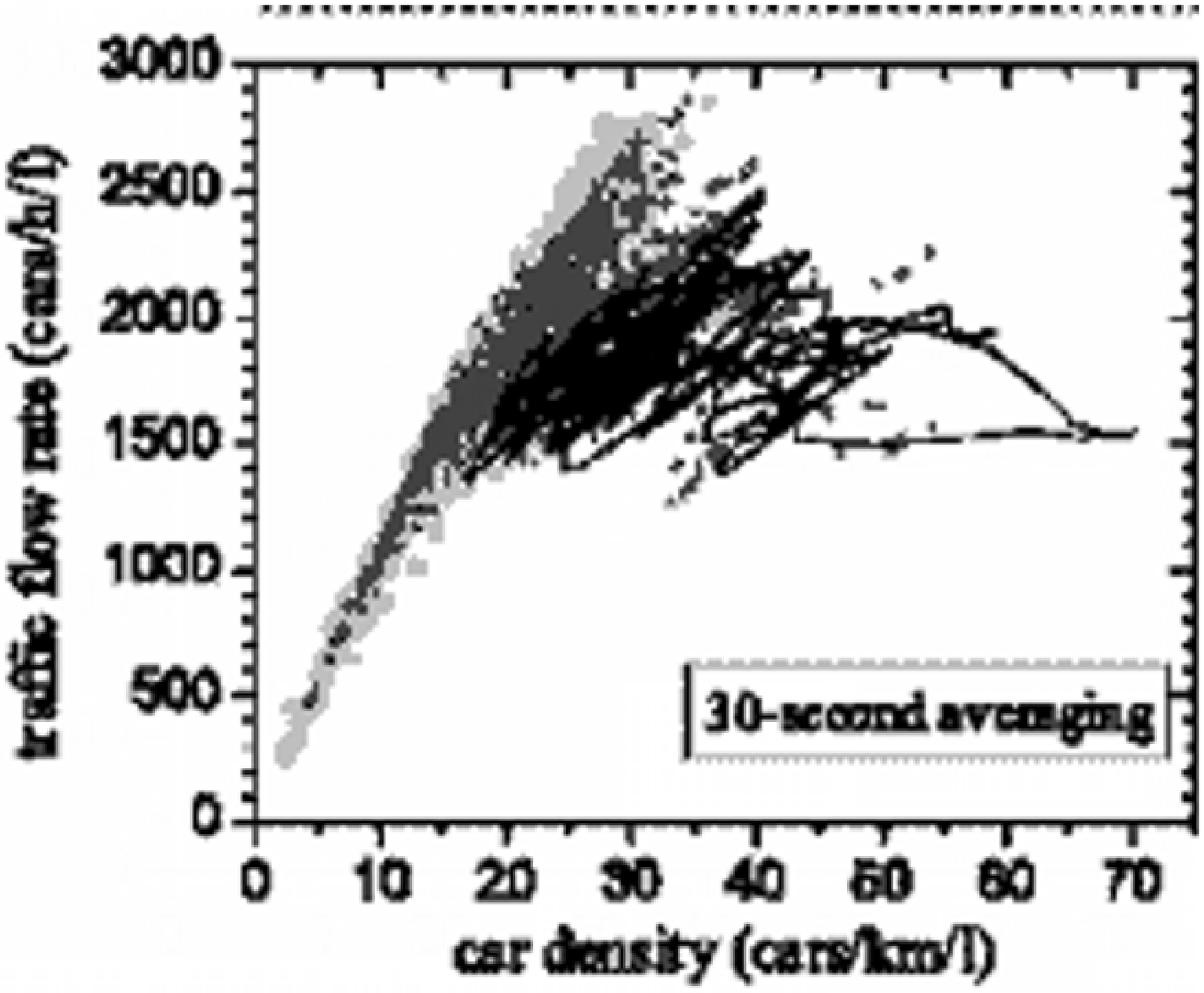,width=0.4\textheight ,height=0.5\textwidth}
\end{center}

\epsfig{figure=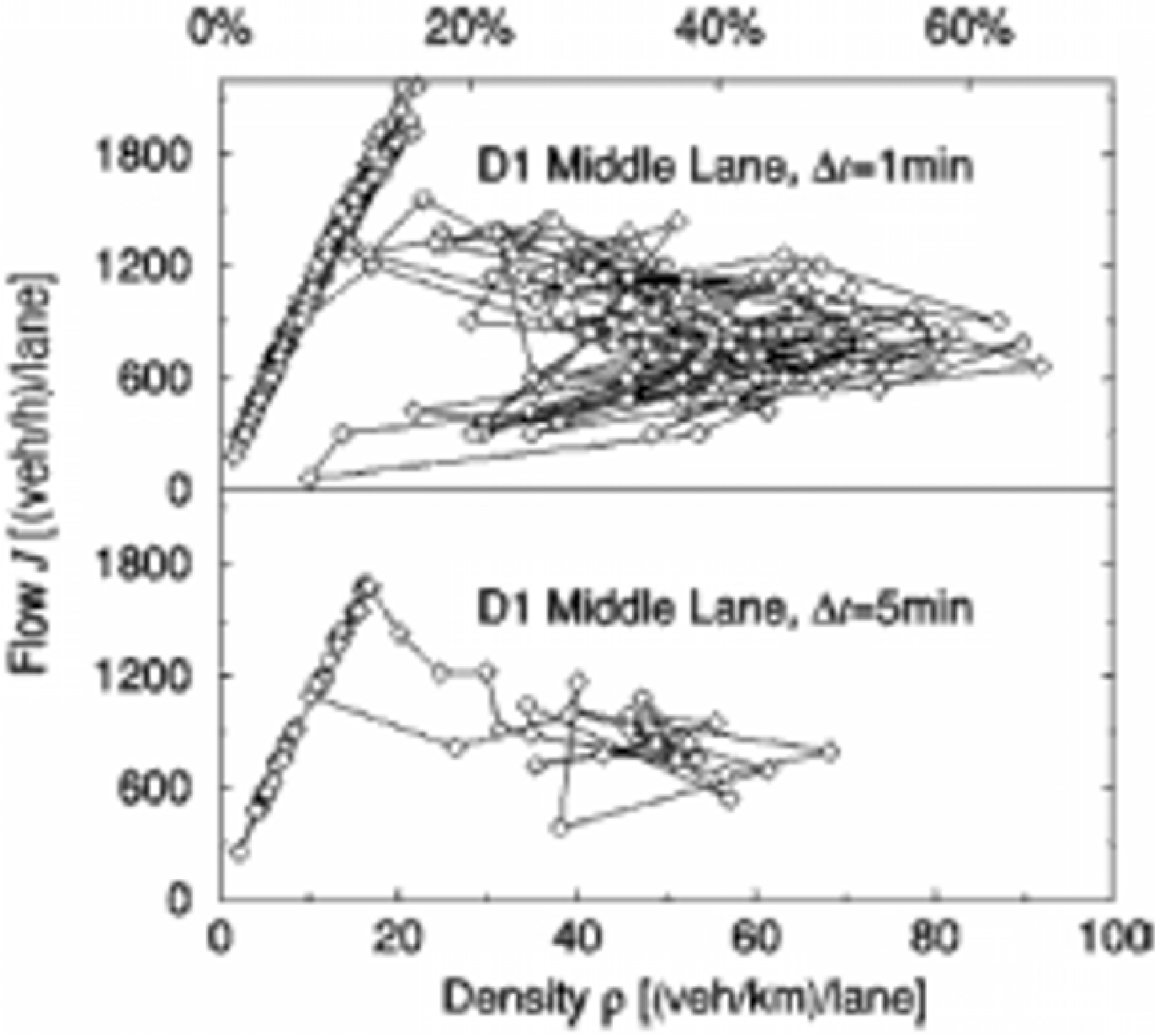,width=0.6\textheight ,height=0.7\textwidth}
\vskip 1cm

\centerline{
Figure 10: Experimental data in field "density--velocity"}
\centerline{
with different intervals of averaging,}
\centerline{
the upper fragment is taken from [4],}
\centerline{the below fragment is taken from [6]
}

\vskip 1cm

The considered model allows to explain the appearance on the
fundamental diagram of scattering domains of density where
different values of intensity can correspond to the same
value of density. The existance of scattered domains has been
discovered on the basis of the experimental results, described in [4,5],
fig. 9--11.

If the flow density $\rho<\rho_{\max}/2$ then the movement is 
stable and the traffic can be described well by a 
hydrodynamic model. In this case the regular movement 
$r=1$ is a single stable state. The effect of random  
disturbances are short-lived and the traffic return fast
to the previous regime.

For sufficiently large densities, that are out of the range  
of the stability of the traffic, the appearances of
disturbances result in that regularity of traffic has
values $r<1$ because of reaction of drivers on changes of the
situation that gives a change of the dynamic dimension.
The density becomes a random value. Dispersion of the
density becomes not equal to zero. The changes of rate
$r/d(v)$ result in that values of densities, which have
been found in the experiments, has dispersion near the
average value when the intensity of the traffic does not
change. Thus the same value of the intensity, obtained
in the experiments, can correspond to differents values of
densities and therefore scattered domains on the
fundamental diagram appear.

\epsfig{figure=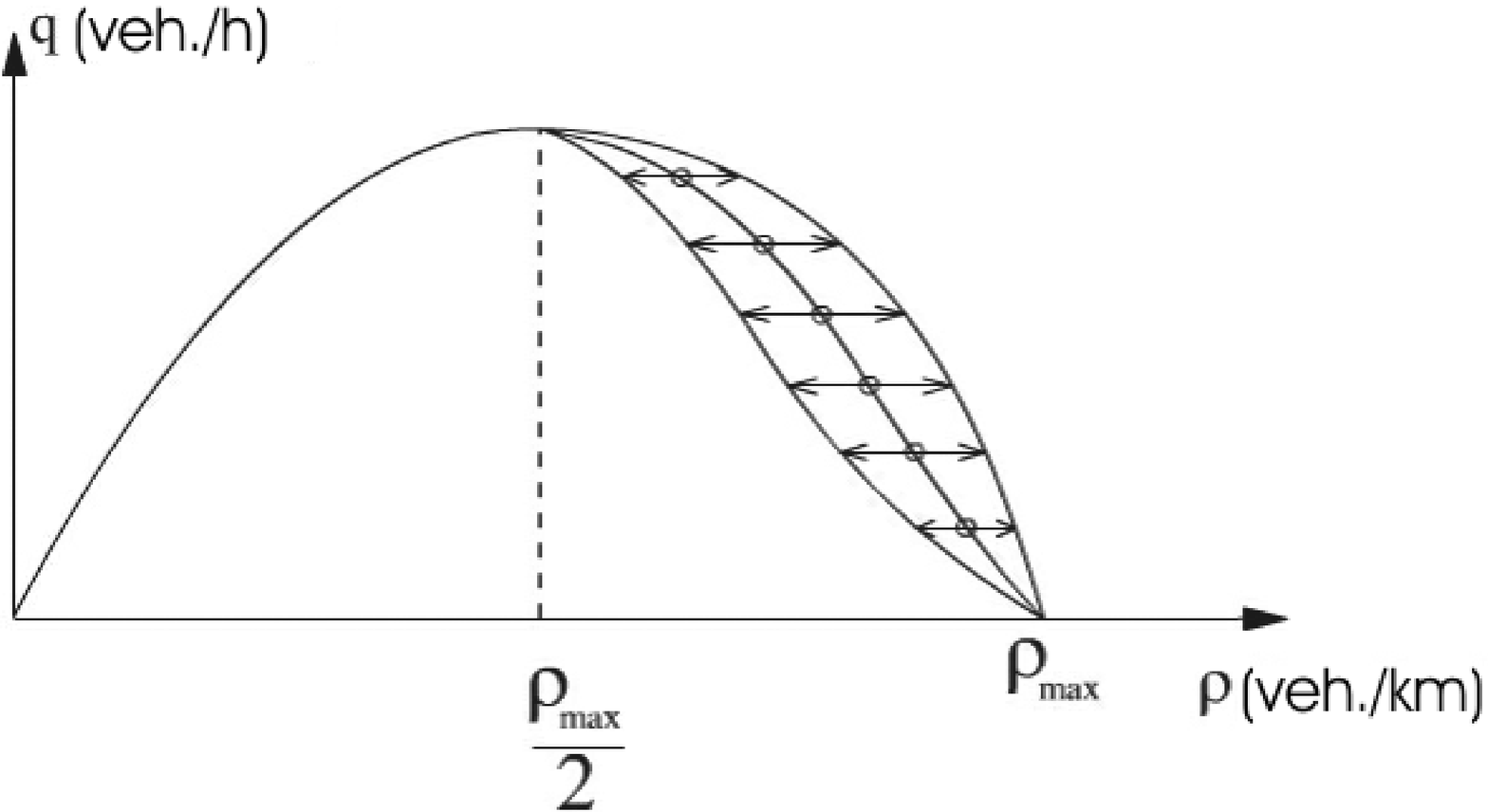,width=0.5\textheight ,height=0.5\textwidth}

\vskip 1cm
\centerline{Figure 11:
Fundamental diagram with scattering domains
}

\vskip 1cm

\subsection*{5. Multilane motion "Regularity -- velocity -- intensity"}

Similarly, for the case of two lanes
$${\bar q}_2={\bar q}_2(r,v,p)=\frac{rv}{c_0+c_1 v+c_2 v^2}+
pr(1-2r^2+r^3).\eqno(11)$$
we have the graphic dependences for the intensity and the velocity
(fig. 12, 13).


\begin{center}
\epsfig{figure=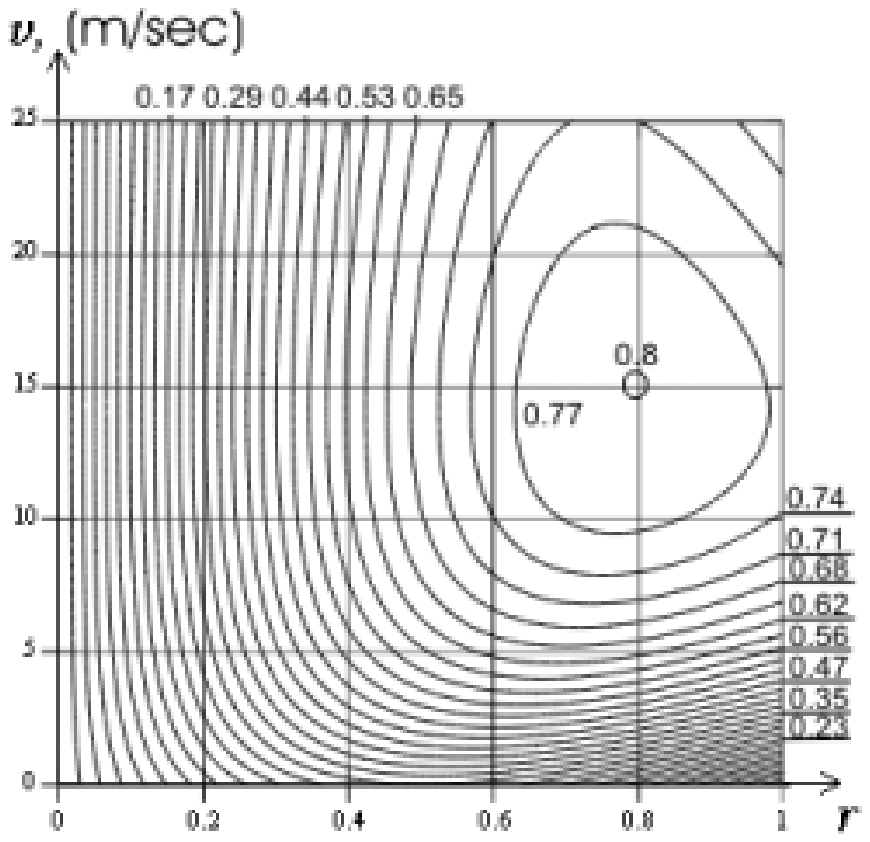,width=.4\textheight ,height=.6\textwidth}
\end{center}

\centerline{Figure 12: Function  $\bar{q}_2 (r,v,1)$ (vehicle/sec)}

\smallskip

\begin{center}
\epsfig{figure=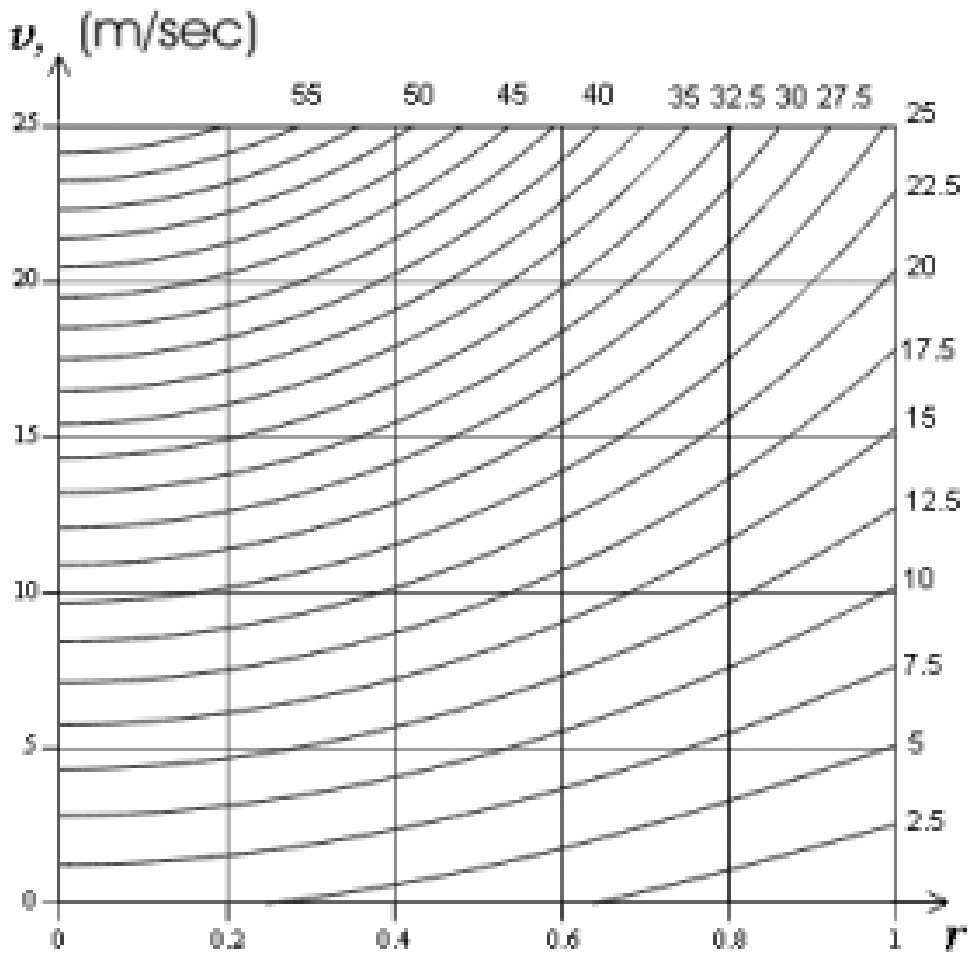,width=.4\textheight ,height=.6\textwidth}
\end{center}

\centerline{Figure 13.  Function  $\bar{v}_2 (r,v,1)$ (m/sec)}


\vskip 1cm

Finally, for three lanes
$${\bar q}_3={\bar q}_3(r,v,p)=\frac{rv}{c_0+c_1v+c_2v^2}+pr(1-4r^3
+4r^4-r^5)\eqno(12)$$
we obtain the dependences shown on fig. 14, 15.


\begin{center}
\epsfig{figure=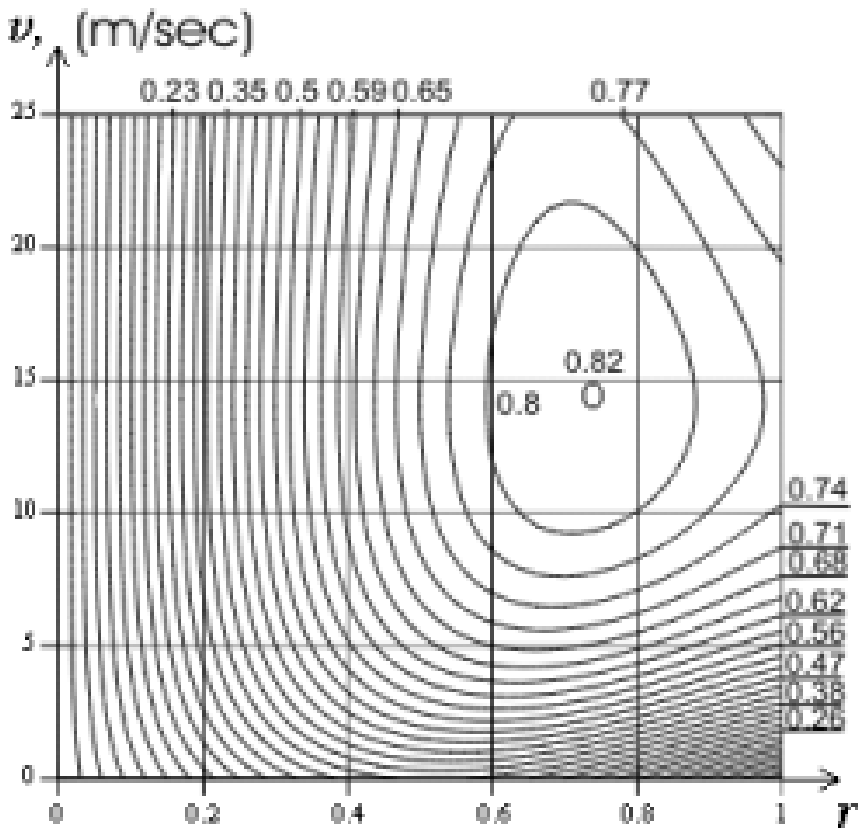,width=.35\textheight ,height=.45\textwidth}
\end{center}

\centerline{Figure 14.  Функция $\bar{q}_3 (r,v,1)$
(vehicle/sec)}

\begin{center}
\epsfig{figure=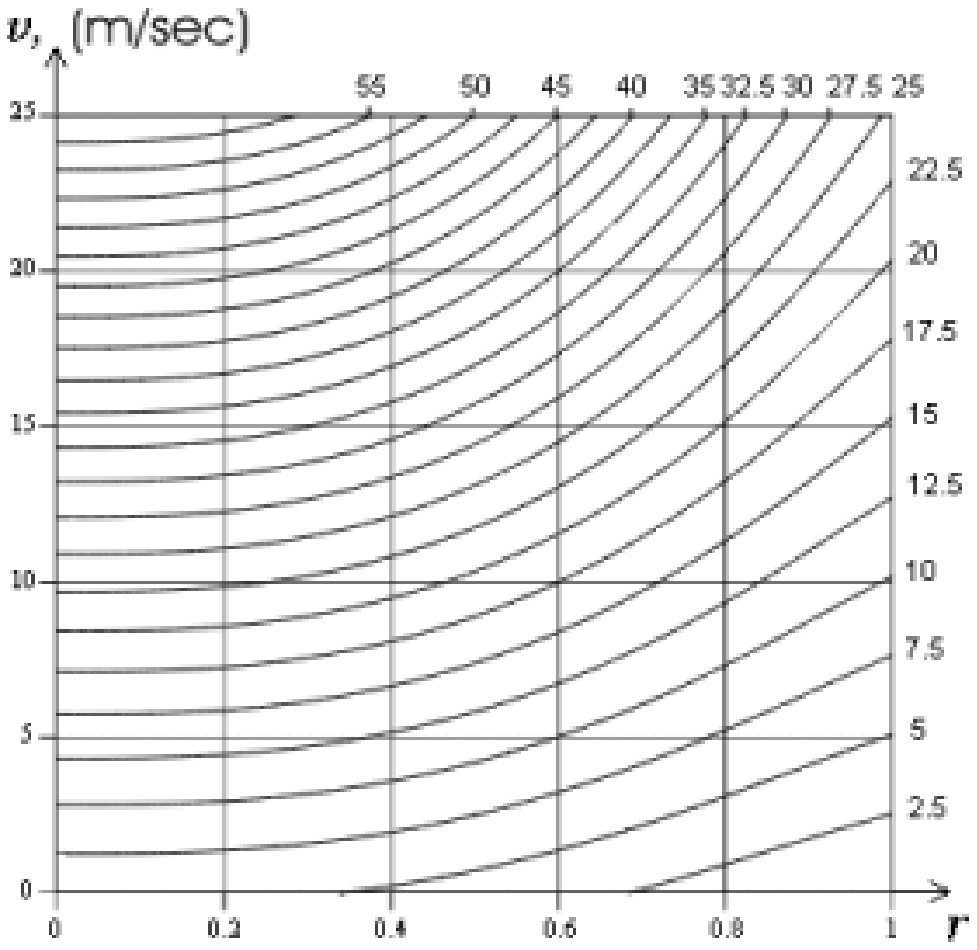,width=.4\textheight  ,height=.6\textwidth}
\end{center}

\centerline{Figure 15.  Функция $\bar{v}_3 (r,v,1)$
(m/sec)}


\vskip 1cm
\subsection*{6. Collective and individual for
the flows with a constant density}

For fixed density $\rho_{0}=r/d(v)$ from (12) we obtain
$$\bar{q}_1={\bar q}_1(\rho,v,p)=$$
$$=v\rho+pd(v)\rho(1-d(v)\rho)=$$
$$=v\rho+p(c_0+c_1v+c_2v^2)\rho(1-\rho(c_0+c_1v+c_2v^2))=$$
$$=d^{-1}(r/v)\rho+pr(1-r).\eqno(13)$$

The unequality $r<1$ is equivalent to unequality $\rho d(v)<1,$
i.e.
$$c_0+c_1 v+c_2 v^2<\frac{1}{\rho},$$
$$v<-\frac{c_1}{2c_2}+\sqrt{(\frac{c_1}{2c_2})^2-\frac{c_0}{c_2}+
\frac{1}{\rho}}.\eqno(14)$$

Let us represent the evaluations of dependence (13) of intensity
on velocity for the fixed density. The question is how the rate changes
if the flow velocity changes abruptly during a small period of time and
the density is constant. If the case of one lane is considered
then for $\rho_{0}=0.01$ vehicle/m 
we have the dependence shown in fig. 16. As for dependences,
shown on fig. 17--19, this dependence corresponds to the case of dry 
asphalt (eq. (2)): $c_0=5.7$ m/sec; $c_1=0.504$ sec; $c_2=
0.0285$ sec/m. Range of velocities below 25 m/sec, i.e 90 km/h,
is considered.


\epsfig{figure=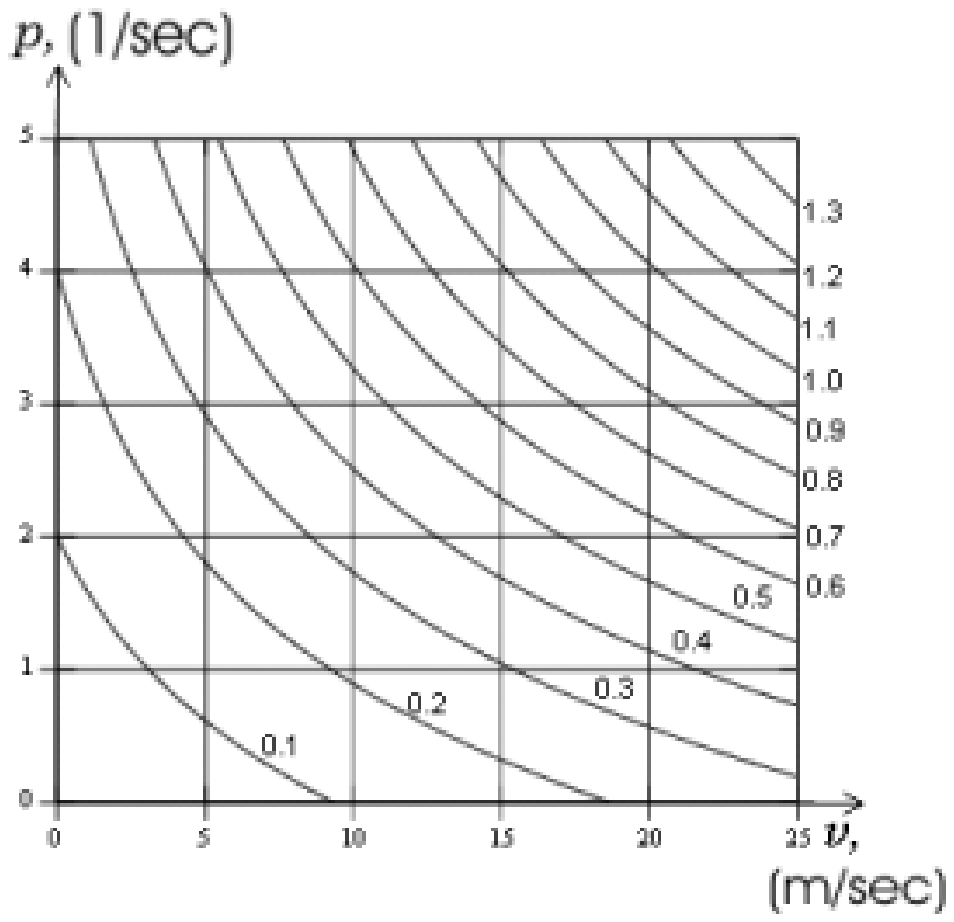,width=.4\textheight  ,height=.6\textwidth}
\smallskip

\centerline{Figure 16.  График $\bar{q}_1(v,p)$ (vehicle/sec),  
$\rho = 0.01 $ (vehicle/m)}
\smallskip

\smallskip
\epsfig{figure=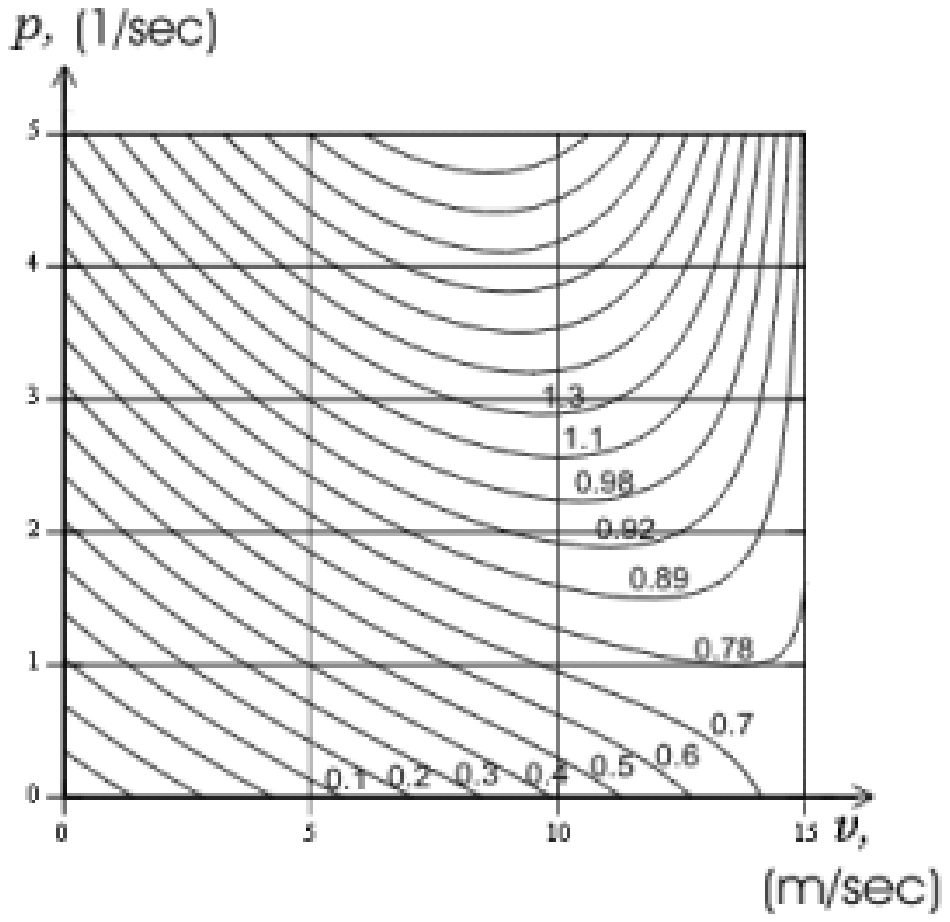,width=.4\textheight  ,height=.6\textwidth}
\smallskip

\centerline{Figure 17.  Dependence $\bar{q}_1(v,p)$ (vehicle/sec), $\rho = 0.05$
(vehicle/m)}

The dependence
for $\rho_{0}=0.05$ is shown on fig. 14.
On fig.17 there is the evident unstability in the neighbourhood of
the right lower angle. 
For the case of the three-lane road we have dependences shown on fig.18-19.


\epsfig{figure=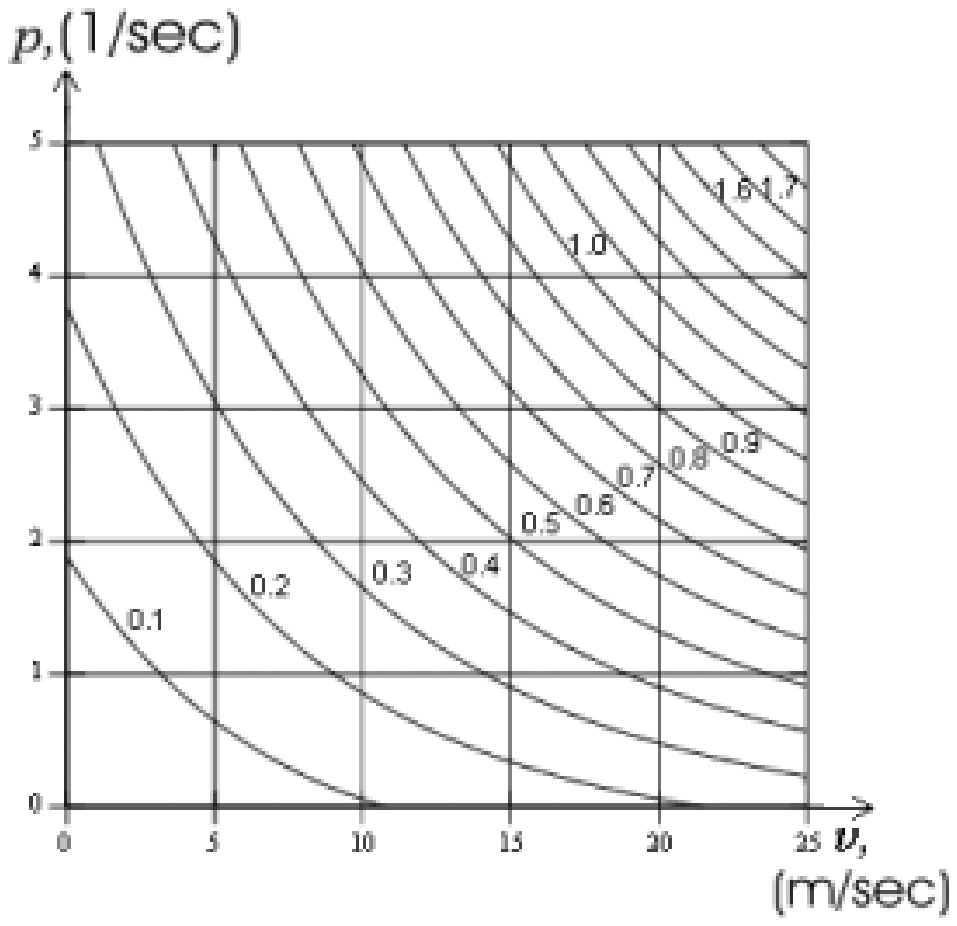,width=.4\textheight  ,height=.6\textwidth}

\centerline{Figure 18. График $\bar{q}_3(v,p)$ (vehicle/sec),
$\rho = 0.01$ (vehicle/m)
}
\epsfig{figure=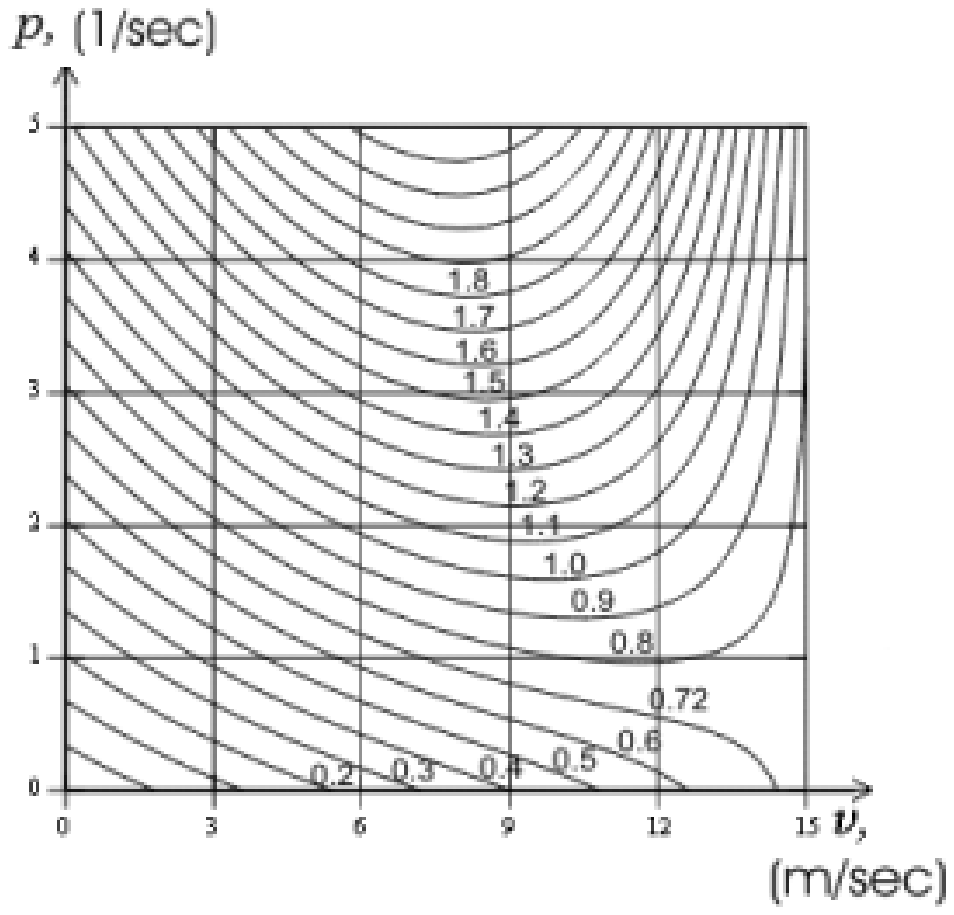,width=.4\textheight  ,height=.6\textwidth}
\smallskip

\centerline{Figure 19.  График $\bar{q}_3(v,p)$ (vehicle/sec),
 $\rho = 0.05 $ (vehicle/m)          
}

\subsection*{7. Influence of "blue lights" upon the flow rate}

The particles of the two types characterized by
of the different functions of dynamic distances
$d_1(v)$ and $d_2(v)$ are considered. Suppose $d_2(v)< d_1(v)$ for each 
allowed $v.$ If $v$ is the regular velocity of the unmixed flow with zero 
individual velocity of the unmixed flow, then by mixing  
between two large particles small particle would emerge 
(percolation). Hence the flow velocity can be evaluated as
$v_1(v)=d_1^{-1}(d_1(v)/(2)).$
Let us suppose still that the number of such particles is rather small. That
the flow intensity in the new conditions is
$$ \rho v_1(v) + \rho 
\frac{p(\rho d(v)/2),T)}{T}(d(v)/2),$$
where the function $p(r,T)$ is defined as the function
$p(r,T)$ introduced in section 1; dynamic distance    
$d(v)$ is calculated according equation (2).

The value $v_1(v),$ which satisfies equation
$d(v_1(v))=\frac{d(v)}{2}$ that is equation 
$$c_0+c_1v_1(v)+c_2v^2_1(v)=\frac{c_0+c_1v+c_2v^2}{2},$$
exists only in the case
$$c_1v+c_2v^2\ge c_0. \eqno(15)$$

Suppose also
$$1-\rho d(v)/2> 0,$$
i.e.
$$c_0+c_1 v+c_2 v^2<\frac{2}{\rho},$$
$$v<-\frac{c_1}{2c_2}+\sqrt{(\frac{c_1}{2c_2})^2-
\frac{c_0}{c_2}+\frac{2}{c_2\rho}}.$$

Thus the equalities
$$
-\frac{c_1}{2c_2}+\sqrt{(\frac{c_1}{2c_2})^2+\frac{c_0}{c_2}}<
v<-\frac{c_1}{2c_2}+\sqrt{(\frac{c_1}{2c_2})^2-\frac{c_0}{c_2}+
\frac{2}{c_2\rho}}.
$$
are true.

\epsfig{figure=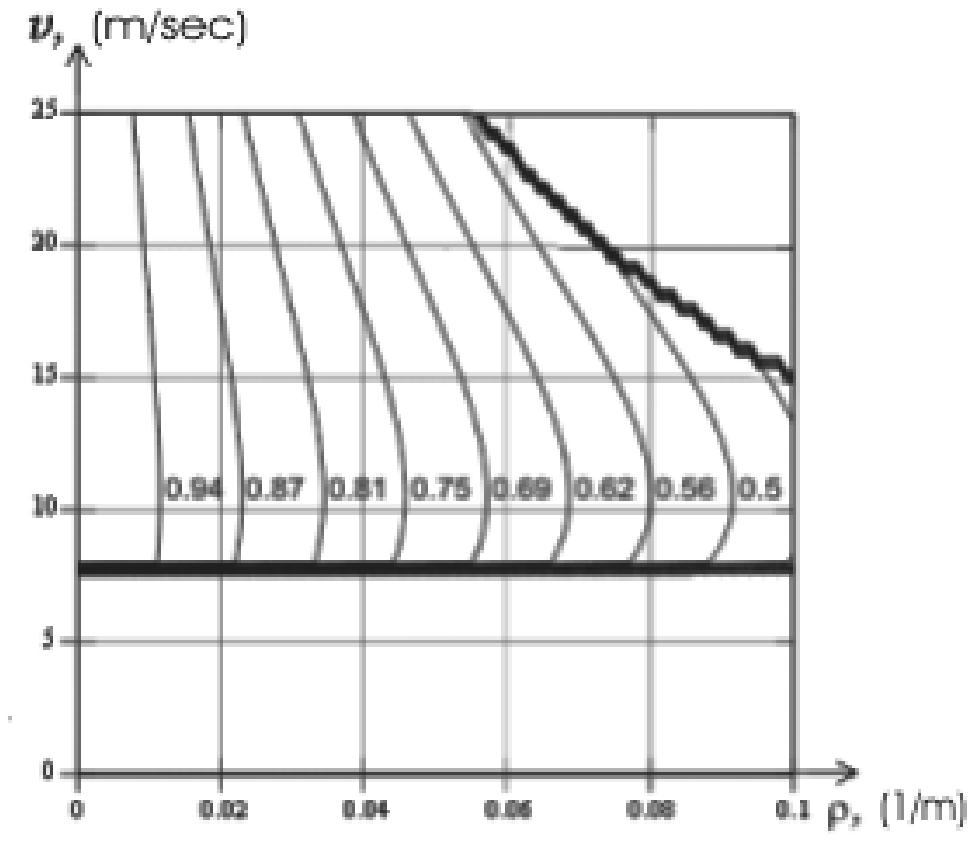,width=.4\textheight  ,height=.6\textwidth}

\centerline{Figure 20. Dependence $Q(\rho,v) $}

Suppose that
$$ p(v)d(v)/2 = v - v_1(v).$$
Hence
$$ q_i(\rho, v)  = \rho \left(v_1(v) + (v - v_1(v)) R_i(\rho d(v)/2)\right).$$

For example, if $i = 1$ we have
$$ q_1(\rho, v)  = \rho (v_1(v) + (v - v_1(v))(1 - \rho d_1(v)/2)).$$
As 
$$ v_1(v) = -\frac{c_1}{2c_2} + \sqrt{ (\frac{c_1}{2c_2})^2 -
(\frac{c_0}{2c_2}- \frac{c_1}{2c_2}v -\frac{v^2}{2})} ,$$
we obtain the ratio of intensity of disturbed and undisturbed conditions
$$  Q(\rho, v) = \frac{v_1(v) + (v- v_1(v))(1- 
\rho d_1(v)/2)}{v} .$$

Relative variation of flow rate of slow particles in case of small number 
of fast special vehicles is shown on fig. 20. We still suppose 
that $c_0=5.7$ m/sec; 
$c_1=0.504$ sec; $c_2=0.0285$ sec/m.

\subsection*{Literature}

\hskip0.6cm 1. Buslaev A.P., Novikov A.N., 
Prikhodko V.M., Tatashev A.G.,  
Yashina~M.V. Stochastic and simulation approaches to
optimisation of traffic. M.: Mir, 2003.--368~c. 

2. Inosse H.., Tanaka T. The road traffic control. 
Moscow: Transport, 1983.

3. Klivkonshtein G.I., Afanasiev M.B. Organization   
of road traffic. Moscow: Transport, 1997. 
1998.--408~с. 

4. Lubashevski I, Mahnke R., Wagner P., Kalenkov S.
Order parameter model unstable multilane traffic// 
Phys. Rev., E 66, 016117 (2002). 

5. Lubashevski I., Wagner P., Mahnke R. Bounded rational
driver models. Eur. Phys. J. B~32, p. 243--247, 2003.

6. Neubert L., Santen L., Schadschneider A., Schreckenberg M.,
Phys. Rev. E 60, 6480 (1999).  

7. Prigogine I., Herman R. Kinetic theory of vehicular traffic//  
American Elsevier, N 5, 1971.

8. K$\ddot{u}$hne R., Mahnke R., Lubashevsky I. Probabilistic
description of traffic breakdowns// Phys. Rev., E 65, 066125, (2002).

\end{document}